\documentclass[a4paper]{jpconf}
\usepackage{graphicx}
\usepackage{amssymb}
\usepackage{epstopdf}
\begin{document}
\title{Quantum statistical calculation of cluster abundances in hot dense matter}

\author{G R\"opke}

\address{Universit\"at Rostock, Institut f\"ur Physik, 18051 Rostock, 
Germany}

\ead{gerd.roepke@uni-rostock.de}

\begin{abstract}
The cluster abundances are calculated from a quantum statistical approach taking into account in-medium corrections. For arbitrary cluster size the self-energy and Pauli blocking shifts are considered. Exploratory calculations are performed for symmetric matter at temperature $T=5$ MeV and baryon density $\varrho=0.0156$ fm$^{-3}$ to be compared with the solar element distribution. It is shown that the abundances of weakly bound nuclei with mass number $4<A<12$ are strongly suppressed due to Pauli blocking effects. 
\end{abstract}

The investigation of thermodynamic properties of hot nuclear matter for parameter values temperature $k_{\rm B}T \approx 1 - 100$ MeV, baryon density $\varrho \leq \varrho_0 \approx 0.166$ fm$^{-3}$ is recently of interest with respect to astrophysical and cosmological questions as well as the investigation of heavy ion collisions. In particular, the problem of the calculation of abundances of clusters $\{AnP\}$ (nuclei with mass number $A$, excitation state $n$ and center of mass momentum $P$) within a systematic quantum statistical approach is of relevance for the equation of state (EoS) and the thermodynamic properties. We will not give here an exhaustive review of the quantum statistical treatment of hot nuclear matter, see \cite{R83,RMS}. We give only some details and results in calculating the cluster abundances, in particular the treatment of heavier clusters at low densities.  

We start from a cluster decomposition of the EoS obtained from a self-consistent ladder Hartree-Fock approximation for the self-energy:
\begin{eqnarray} \label{1}
&&\varrho (\beta, \mu) = \sum_{A=1}^\infty A \varrho_A(\beta, \mu)=  \sum_{A=1}^\infty A \sum_n^{\rm bound} g_{A,n} \int_{P > P^{\rm Mott}_{A,n}} \frac{d^3P}{(2 \pi)^3} f_A(E_{AnP}),\nonumber \\ 
&& f_A(E)=\left[\exp \beta (E-A \mu)- (-1)^A \right]^{-1},
\end{eqnarray}
where $\varrho_A(\beta, \mu)$ denotes the contribution of clusters $A$ to the total baryon density, $g_{A,n}$ is the degeneration factor of the state $n$. The energy eigenvalue $E_{AnP}  $ of the cluster $\{AnP\}$ is obtained from a Bethe-Goldstone equation that takes into account the influence of the surrounding matter on the quantum state of the considered cluster. In ladder Hartree-Fock approximation it can be decomposed as follows: 
\begin{equation}
E_{AnP}=E_{AnP}^0(\varrho, T)+\frac{P^2}{2 A M}+ \Delta E_{AnP}^{\rm SE}+ \Delta E_{AnP}^{\rm Pauli},
\end{equation}
where $E_{AnP}^0(\varrho, T)$ describes the dependence of the energy of the cluster because of the Coulomb interaction with the surrounding charged particles, $\Delta E_{AnP}^{\rm SE}$ is the self-energy shift, and $\Delta E_{AnP}^{\rm Pauli}$ denotes the shift due to phase space occupation (Pauli blocking). The summation in Eq. (\ref{1}) concerns only the bound states. This means that the condition
\begin{equation} \label{3}
E_{AnP}<A \Delta(P/A)
\end{equation}
holds, since the lower edge of the continuum of scattering states for each of the A cluster constituents is shifted by the single particle self-energy shift $\Delta(p)$ at momentum $p=P/A$. Above the so-called Mott density introduced in \cite{RMS}, the bound state merges in the continuum of scattering states and disappears. The condition (\ref{3}) excludes the interval $P <P^{\rm Mott}_{A,n}(\varrho,T)$ for the integral over $P$ in Eq. (\ref{1}) if the density exceeds the Mott density.

To evaluate the EoS (\ref{1}) we have to determine the energies $E_{AnP}$. We give some details considering as example the parameter values $T= 5$ MeV and $\varrho = 0.0156$ fm$^{-3}$. In addition to the ladder-Hartree Fock approximation and the approximation for the evaluation of the Pauli shift given below that the surrounding of the cluster under consideration is described as a nondegenerated, uncorrelated gas of nucleons, we assume symmetric nuclear matter where protons and neutrons are considered as particles interacting in the same manner. Within the exploratory model calculations given here for symmetric matter, we make the approximation that we attribute to a system of $A$ nucleons the charge $eA/2$, this means, each nucleon (mass $M$) has the charge $e/2$.  For symmetric matter, we have only one chemical potential $\mu$, for asymmetric matter see \cite{R83} where the two-component case is treated. Further assumptions  refer to empirical quantities (radii of nuclei, density distribution within a cluster, single-particle Hartree-Fock shift, density of states of excited nuclei, and so on) that can be improved in a systematic way. Our main interest is the calculation of the energy shift of heavier elements, in particular $4<A<12$. 

The relevance of Coulomb interaction is increasing with increasing cluster size $A$ and leads to two effects: The nucleon density near a cluster is reduced due to the Debye-Thomas-Fermi theory of screening of a charge distribution by the surrounding plasma, whereas the density of the strongly degenerated electrons (as compensating background) remains nearly constant. At the surface of the cluster $A$, given by the radius $R_A=r_0 A^{1/3}$ with $r_0=1.2$ fm, we have the density $\varrho_A^*(\varrho,T)<\varrho$. Furthermore, the energy of a cluster is shifted to the value $E^0_{AnP}(\varrho,T)$ due to the screening of the Coulomb potential.

The screening radius $1/\kappa$ is given by
\begin{equation}
\label{4}
\kappa^2=\pi e^2 \frac{d}{d \mu} \varrho(\mu),
\end{equation}
where the value $\mu$ in eq. (\ref{4}) is obtained from the relation for the ideal Fermi gas ($\Lambda^2 = 2 \pi \hbar^2/M k_BT$)
\begin{equation}
\label{5}
\varrho(\mu)=\frac{2}{\pi^2 \Lambda^3} \int dx \frac{x^2}{\exp (x^2/4 \pi - \mu/k_BT)+1}.
\end{equation}
For the parameter set given above, $\mu/k_BT=0.892;\,\,\, k_BT/\varrho \times d\varrho/d \mu=0.668$  results.
According to the Debye theory follows 
 \begin{equation}
\label{6}
\varrho_A^*(\varrho,T)=\varrho \exp\left\{-\frac{\kappa^2 A}{4 \pi \varrho R_A}\frac{1+\gamma}{1+\kappa R_A}\right\}.
\end{equation}
\begin{eqnarray}
\label{7}
&&E^0_{AnP}(\varrho,T)-E^0_{AnP}(0)=\frac{e^2}{4 r_0} A^{5/3}\left\{\frac{3}{5} (1+\gamma)^2-\frac{3}{5} \right.\nonumber \\
&&\left.-\frac{1}{2}(1+\gamma)^2\frac{\kappa R_A}{(1+\kappa R_A)^2}\left( \frac{3}{2} +\kappa R_A\right) \right\}.
\end{eqnarray}
 \begin{equation}
\label{8}
\gamma=\frac{\varrho}{\varrho_0}\left[\exp \left\{ -\frac{e^2}{4} \frac{A}{R_A k_BT}(1+\gamma) \left(\frac{1}{1+\kappa R_A} +\frac{1}{2} \right) \right\}-1\right].
\end{equation}
Results for different values $A$ are given in Tabs. 1,2 of Ref. \cite{R83} where for $E_{AnP}^0(0)$ the value of the empirical binding energy for the most abundant isobar is taken; the dependence of the energy $E^0_{AnP}(\varrho,T)$ from the center of mass momentum $P$ has been neglected.

For the parameter values considered here, the values $\varrho_A^*(\varrho,T)$ for $A\leq 11$ exceeds the Mott density so that for these cluster the value $P^{\rm Mott}_A \neq 0$ has to be determined where the cluster breaks up into free nucleons or smaller clusters (with energies shifted according to the density $\varrho^*_A(\varrho,T)$, Eq. (\ref{6}), as shown below). This way, the interval of values $P$ for that the cluster can exist, is reduced.

The single-particle self-energy shift $\Delta (\varrho)$ can be determined from a Skyrme interaction with zero range so that it is not dependent of temperature and momentum. According to \cite{2} we have (units MeV, fm)
 \begin{equation}
\label{9}
\Delta (\varrho)=-792.975 \varrho +2711.9 \varrho^2+125.226 \varrho^{5/3}.
\end{equation}
For light clusters $A<12$, the self-energy shift is taken as $\Delta E_{AnP}^{\rm SE}= A \Delta(\varrho^*_A)$. For  $A \geq 12$ we use within the Fermi gas model 
 \begin{equation}
\label{10}
\Delta E_{AnP}^{\rm SE}=\int d^3r \Delta(\varrho_A(r)) \rho^*_A,
\end{equation}
where $\rho_A(r)$ is the nucleon density as function of the radius $r$ for the cluster $A$. According to \cite{3} we can use the density distribution
 \begin{equation}
\label{11}
\rho_A(r)=\frac{3A}{4 \pi R^3} \frac{1}{1+(\pi b/R)^2} \left[\frac{1}{1+e^{\frac{r-R}{b}}}+\frac{1}{1+e^{\frac{-r-R}{b}}}-1\right]
\end{equation}
with parameter values $b=0.57$ fm, $R=1.05$ fm $A^{1/3}$.

The Pauli shift has been determined for light clusters $(A\leq 12)$ approximating the cluster wave function by an antisymmetrized product  of single-particle wave functions $\varphi_\nu(i)$:
 \begin{equation}
\label{12}
\Delta E_{AnP}^{\rm Pauli}=-\sum_\nu \sum_1 |\varphi_\nu(1)|^2  \left(E_\nu-\frac{\hbar^2 p_1^2}{2 M} \right) \frac{ \rho^*_A \Lambda^3}{4} {\rm e}^{-\frac{\hbar^2(p_1-P/A)^2}{2 M k_B T}}.
\end{equation}
This expression is evaluated for $4 <A<12$ using the eigenstates of the harmonic oscillator according to Ref. \cite{4} with potential $V(r) = V_0 +M\omega^2r^2/2,\,\,\, \hbar \omega = 40 A^{-1/3}$ MeV. The single-particle Hartree-Fock energies $E_\nu$ and the the binding energy $E_{A0P}^0(\varrho,T)$ lead to the relation
 \begin{equation}
\label{13}
V_0=\frac{2}{A} \left(E_{A0P}^0(\varrho,T)+3 \hbar \omega\right)- \frac{15}{4} \hbar \omega.
\end{equation}

For $A\leq 4$ we also can use a product of Gauss functions $\varphi_\nu(p) \propto \exp(-p^2/2 k_A^2)$ with the parameter
 \begin{equation}
\label{14}
 k_A^2=4 M |E^0_{A0P}|/A \hbar^2.
\end{equation}
As before, we find the relation 
 \begin{equation}
\label{15}
E_\nu=2 E^0_{A0P}(\varrho,T)/A-3 \hbar^2 k_A^2/4 M= 5 E^0_{A0P}(\varrho,T)/A.
\end{equation}
The Pauli blocking shifts for $A \leq 4$ follow from Eq. (\ref{12}) as
\begin{eqnarray}
\label{16}
&&\Delta E_{A0P}^{\rm Pauli}=A \frac{\varrho_A^* \Lambda^3}{4} \frac{|E^0_{A0P}|}{\delta^{3/2}} \left\{5+\frac{3}{\delta}+\frac{x^2(\delta-1)}{2 \pi \delta^2}\right\}e^{-\frac{x^2}{4 \pi \delta}} \nonumber \\
&& \delta=1+2 |E^0_{A0P}|/AT,\qquad x=P \Lambda/A.
\end{eqnarray}
For $4 < A < 12$ we have according Eq. (\ref{12})
\begin{eqnarray}
\label{17}
&&\Delta E_{A0P}^{\rm Pauli}= \frac{\varrho_A^* \Lambda^3}{4} e^{-\frac{x^2}{4 \pi \delta}}
\frac{\hbar \omega}{\delta^{3/2}} \left\{\frac{3}{\delta}-4 \left( \frac{V_0}{\hbar \omega}+\frac{3}{2}  \right)   +\frac{x^2(\delta-1) \hbar \omega}{2 \pi \delta^2}\right\} \nonumber \\
&& -2(A-4)\frac{1 }{3 \delta}\left[\left( \frac{V_0}{\hbar \omega}+\frac{5}{2}  \right)-\frac{x^2(\delta-1) }{8 \pi \delta^2} \right]\nonumber \\
&& -2(A-4)\frac{x^2(\delta-1) }{12 \pi \delta^2}\left[-\frac{5x^2(\delta-1) }{8 \pi \delta^2} +\frac{3}{2}\left( \frac{V_0}{\hbar \omega}+\frac{5}{2}  \right) + 2(A-4)\frac{5 }{8 \delta^2} \right]\nonumber \\
&& \delta=1+\frac{\hbar \omega}{2 k_BT}.
\end{eqnarray}

For large cluster, $A\leq 12$, the Pauli blocking shift is calculated within a Fermi gas model so that the compensation between self-energy shift and Pauli blocking shift at low temperatures is correctly reproduced. The total shift results as
\begin{eqnarray}
\label{18}
&&\Delta E_{A0P}=\varrho_A^*\int d^3r \Delta(\varrho_A(r))\int_{ \Lambda p_F(\varrho_A)}^\infty \frac{y dy A}{2 \pi P \Lambda} \left\{e^{-(y-P \Lambda/A)^2/4 \pi}-e^{-(y+P \Lambda/A)^2/4 \pi}\right\},
\nonumber \\
&& p_F(\varrho_A)=\left(\frac{3 \pi^2}{2} \varrho\right)^{1/3}.
\end{eqnarray}
The integral over $y$ depends on $\varrho_A(r)$ an can be approximated by $\exp(-\delta \varrho^{2/3})$. To fit the value at $\varrho_0$, the value $\delta=20.926$ results for $P \Lambda=0.3$.

The sum over the excited states $n$ was performed using the density of states \cite{4}
\begin{equation}
D_A(E)=\frac{1}{12}\left(\frac{\pi^2}{a}\right)^{1/4} E^{-5/4} \exp\left(2 \sqrt{a E}\right)
\end{equation}
with $a = A/15$ MeV$^{-1}$. This way, the abundance
\begin{eqnarray}
&&X_A=\varrho_a/\varrho=\frac{A^3}{2 \pi^2 \varrho \lambda^3} \int dx x^2  \int dE D_A(E) \nonumber\\
&&\exp\left\{-(E^0_{A0P}(\varrho,T)+E+\Delta E_{A0P}(x,\varrho_A^*)+ATx^2/4 \pi-A\mu)/k_BT\right\}
\end{eqnarray}
results for cluster with $A\geq 12$. It has been assumed that the shifts are independent of $n$. The evaluation of $X_A$ is simplified if the shift $\Delta E_{AnP}$ is taken for a representative value $x=P \Lambda/A=0.3$ considering the dependence on $P$.

For the light cluster $A<12$, the density considered here exceeds the Mott density. We take into account only the ground state with degeneration factors $g_{2,0}=3, g_{3,0}=4, g_{A,0}=1$ else, calculating the abundances $X_A$. The result for the cluster abundances for the parameter values  $T= 5$ MeV and $\varrho = 0.0156$ fm$^{-3}$ exemplarily considered here are shown in Fig. 1. These parameter values have been chosen to reproduce the global behavior of the solar element distribution. The results  are of interest considering the evaporation of matter from the surface of a dense hot nuclear system, see \cite{R82}.

\begin{figure}[ht] 
	\includegraphics[width=0.8\textwidth]{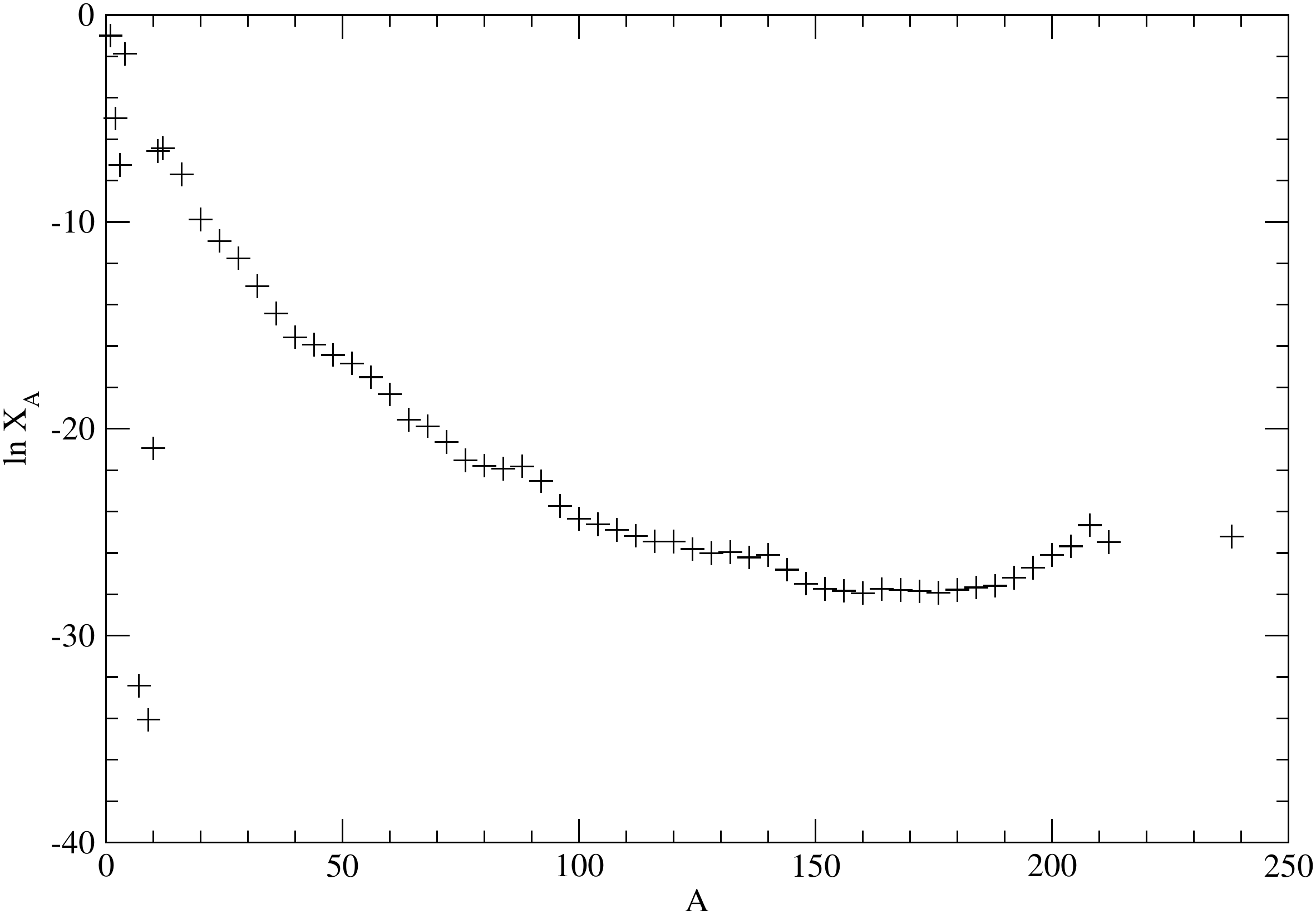}
	\caption{Cluster abundances  $\ln X_A$ for hot and dense symmetric matter as function of mass number $A$, parameter values $T= 5$ MeV and $\varrho = 0.0156$ fm$^{-3}$. Most abundant isobars are considered for $A<12$. For $A\geq 12$ abundances of clusters with $A=4n$ are shown.}
 \label{Fig:1} 
 \end{figure}

The approach given here considers in particular for the first time the abundance of the light clusters $4<A<12$ that are weakly bound so that they are strongly suppressed in hot dense matter. Some approximations can be improved. The inclusion of scattering states in Eq. (\ref{1}) is considered in the cluster-virial expansion of the EoS \cite{R13}. A more detailed calculation of the Pauli blocking shift for the light clusters $A \leq 4$ is given in \cite{R2011}, and the EoS that contains only these light clusters was discussed in Ref. \cite{unsere}. The inclusion of arbitrary clusters calculating the nuclear matter EoS was considered in \cite{Hempel2010} where the medium effects are taken into account using the phenomenological excluded volume concept. The strong influence of medium effects on the abundance of elements with  $4<A<12$ is caused by the week binding of these cluster,
similar to the deuteron that is also strongly influenced by medium effects.


\section*{References}

\begin{thebibliography}{99}

\bibitem{R83}
  R\"opke G 1983 \textit{Phys. Lett. B} {\bf 121} 223;
  R\"opke G 1984 \textit{Wiss. Z. Univ. Rostock} {\bf 33} 33
  
\bibitem{RMS}
  R\"opke G, M\"unchow L, Schmidt M and Schulz H
  1982 \textit{Nucl. Phys. A} {\bf 379} 536;
  1983 \textit{Nucl. Phys. A} {\bf 399} 587;
  1984 \textit{Nucl. Phys. A} {\bf 424} 594;
  1982 \textit{Phys. Lett.} {\bf B 110} 21
  
\bibitem{R82}
   R\"opke G 1983 \textit{Wiss. Z. Univ. Rostock} {\bf 32} 30

\bibitem{2} 
Vautherin D and Brink D M 1972 \textit{Phys. Rev. C} {\bf 5} 626 
 
\bibitem{3} 
Burov V V, Eldyshev Yu N, Lukyanov V K and Pol Yu S 1974 \textit{JINR preprint} E4-8029 Dubna 
 

\bibitem{4} 
Bohr A and Mottelson B 1963 {\it Nuclear Structure}, Benjamin, New York 

\bibitem{R13} 
R\"opke G, Bastian N-U, Blaschke D, Kl\"ahn T, Typel S and
Wolter H H 2013 \textit{Nucl. Phys. A} {\bf 897} 70
 
\bibitem{R2011}
  R\"opke G 
  2011 \textit{Nucl. Phys. A} {\bf 867} 66
 
 \bibitem{unsere}
	Typel S, R\"opke G, Kl\"ahn T, Blaschke D and Wolter H H, 
	2010 \textit{Phys. Rev. C} {\bf 81} 015803 
 
\bibitem{Hempel2010}
	Hempel M and Schaffner-Bielich J
% 	{\it A statistical model for a complete supernova equation of state},
	2010 \textit{Nucl. Phys. A} {\bf 837} 210

\end{thebibliography}
\end{document}